\documentclass[runningheads]{llncs}
\usepackage{graphicx}
\usepackage{subfig}
\usepackage{hyperref}
\usepackage{tabularx}
\usepackage{longtable}
\usepackage{tikz,xcolor,hyperref}

\definecolor{lime}{HTML}{A6CE39}
\DeclareRobustCommand{\orcidicon}{%
	\begin{tikzpicture}
	\draw[lime, fill=lime] (0,0) 
	circle [radius=0.16] 
	node[white] {{\fontfamily{qag}\selectfont \tiny ID}};
	\draw[white, fill=white] (-0.0625,0.095) 
	circle [radius=0.007];
	\end{tikzpicture}
	\hspace{-2mm}
}

\foreach \x in {A, ..., Z}{%
	\expandafter\xdef\csname orcid\x\endcsname{\noexpand\href{https://orcid.org/\csname orcidauthor\x\endcsname}{\noexpand\orcidicon}}
}

\begin{document}
\title{Coordinating Narratives and the Capitol Riots on Parler \thanks{The research for this paper was supported in part by the Knight Foundation and the Office of Naval Research Grant (N000141812106) and an Omar N. Bradley Fellowship, and by the center for Informed Democracy and Social-cybersecurity  (IDeaS) and the center for Computational Analysis of Social and Organizational Systems (CASOS) at Carnegie Mellon University. The views and conclusions  are those of the authors and should not be interpreted as representing the official  policies, either expressed or implied, of the Knight Foundation, Office of Naval Research or the US Government.}}

\author{ Lynnette Hui Xian Ng\inst{1}\orcidA \and Iain Cruickshank\inst{1}\orcidB \and 
Kathleen M. Carley\inst{1}\orcidC}
\authorrunning{Ng et. al.}
\titlerunning{Uncovering Coordination behind Parler's Capitol Riots}

\institute{CASOS, Institute for Software Research \\
Carnegie Mellon University, Pittsburgh, PA 15213 \\
\email{\{huixiann, icruicks, carley\}@andrew.cmu.edu}}

\maketitle              

\begin{abstract}
Coordinated disinformation campaigns are used to influence social media users, potentially leading to offline violence. In this study, we introduce a general methodology to uncover coordinated messaging through analysis of user parleys on Parler. The proposed method constructs a user-to-user coordination network graph induced by a user-to-text graph and a text-to-text similarity graph. The text-to-text graph is constructed based on the textual similarity of Parler posts. We study three influential groups of users in the 6 January 2020 Capitol riots and detect networks of coordinated user clusters that are all posting similar textual content in support of different disinformation narratives related to the U.S. 2020 elections.

\end{abstract}

\section{Introduction}
The free speech social network Parler has been under the spotlight for its role as a preparatory medium for the 6 January 2020 Capitol riots, where supporters of then-president Donald Trump stormed the Capitol building. 
The United States have previously seen protests like the BlackLivesMatter and the StopAsianHateMovements fuelled by disinformation narratives on social media, which can lead to offline violence \cite{zannettou2019disinformation,lyu2021state}. 

A key part of the field of social cybersecurity is the detection of disinformation on social media platforms. Disinformation, in this context is a deliberately manipulated narrative or distortion of facts, to serve a particular purpose \cite{Ireton:2018}. In particular, we are interested in the coordination of spreading disinformation by online actors; coordinated operations can pose threats to the social fabric \cite{carley2020social}. Within the realm of coordinated disinformation campaigns, we aim to identify key actors that spread harmful narratives. In particular, this study seeks to adress which actors, or classes of actors, are involved in disinformation spread on Parler? For this, we analyzed the activity of the users on Parler, with a special focus on three account types: military, patriot and QAnon accounts. We then combined textual and user-level information to analyze coordinated messaging and find influential accounts amon those identified classes of users. 

The main contribution of our work are as follows:
\begin{enumerate}
    \item We identify three classes of accounts, namely those that identify as military or ``patriot" or QAnon that are very influential on Parler after the Januray 6th Capitol riot and active in perpetrating disinformation themes.
    \item We propose a new technique to find coordinated messaging among users using features of their parleys. This text-coordination technique reduces complex networks into core members, providing leads for further analysis.
\end{enumerate}

\section{Background}
Parler positions itself as a free speech social network, where its users can ``speak freely and express yourself openly". A significant portion of its user base are Donald Trump supporters, conservatives, QAnon conspiracy theorists and far-right extremists \cite{aliapoulios2021early}. 
Additionally, the platform has been in the spot light for being a coordination platform for the January 6 2021 Capitol riots \cite{Munn_2021}, where hundreds of individuals stormed the US Capitol Hill building, emboldened by calls from then-president Donald Trump to stop the certification the 2020 U.S. election. The social network was shut down on January 11.

Prior work analyzed the Parler's role in shaping the narratives of different contexts. 
Parleys between August 2018 and January 2021 revealed a huge proportion of hashtags supporting Trump on the elections \cite{aliapoulios2021early}.
Analysis of the platform on the COVID-19 vaccine disproportionately slants towards anti-vaccine opinions \cite{baines2021scamdemic}.
Specific to the capitol riots, a comparative study between Twitter and Parler shows that a huge proportion of Parler activity was in support of undermining the validity of the 2020 US Presidential elections, while the Twitter conversation disapproved Trump \cite{hitkul2021capitol}.

One point of concern is the detection of coordinated campaigns used to influence users, which may therefore lead to real-world violence like the Capitol riot. Several techniques of coordination detection on social media have been previously researched: using network community detection algorithms \cite{nizzoli2020coordinated}, statistical analysis of retweets, co-tweets, hashtags and other variables \cite{vargas2020detection}, and constructing a synchronized action framework
These works, however, depend largely on a network-based structure, which is not dominant in Parler. Another way of assesing coordinated behavior between social media websites is comparing the textual content that users are sharing. Pachecho et al. have recently utilized a textually-based user comparison as a means of overcoming limitations in finding coordinated user behavior with just explicit user-to-user networks \cite{Pacheco:2020}.

\section{Data and Methodology}

In this section, we detail the data used in our study along with the analysis techniques for understanding the actors who spread content on Parler and the narratives present within that content. We first detail the data used and how it was processed for analysis. We then detail how we analyzed the textual content for disinformation themes and coordinating propagation of disinformation.

\subsection{Data Collection and Processing}

The data used for this study comes from the Parler data leak following the January 6th Capitol Riot. After the Capitol riots, Amazon Web Services banned Parler from being hosted on its website. Shortly before that, an independent researcher and other internet users sought to preserve the data by performing a scrape of its posts, or ``parleys" \cite{lyons_2021}. In particular, we use a partial HTML scrape of the posts\footnote{\url{https://ddosecrets.com/wiki/Parler}}. The data set consists of ~1.7 million posts from ~290,000 unique users and has 98\% of its posts from January 3rd to January 10th of 2021. 

We developed a parsing tool to parse the useful elements from the HTML: users, text, external website URLs, echo counts, impression counts and so on \footnote{Code for parsing Parler HTML is available at: \url{https://github.com/ijcruic/Parse-Parler-Data}}. Table \ref{tab:parlerterms} presents some terminology used by the Parler platform.

\begin{table}[]
\begin{tabular}{|l|p{10cm}|}
\hline
\textbf{Term}   & \textbf{Definition}   \\ \hline
Parley & A post on the Parler platform, up to a length of 1000 text characters \\ \hline
Echo & A re-posting of a Parley, similar to Twitter's re-tweet feature\\ \hline 
Hashtag & A word or phrase prefixed with the pound (\#) sign, usually identifying topics in the text \\ \hline 
Comment & A reply to a Parley, up to a length of 1000 text characters \\ \hline 
Upvotes & An indication of approval of the Parley, similar to Twitter's ``like" feature\\ \hline 
Impressions & Appearance of the Parley on another user's feed\\ \hline 
\end{tabular}
\caption{Terminology used in Parler}
\label{tab:parlerterms}
\vspace{-1.3cm}
\end{table}

\subsection{User Class Analysis} 
For the users, we wanted to focus on those users that are most likely to be involved with disinformation or who had an outsize influence in the January 6th Capitol Riot. In particular, we coded for three classes of users based upon news of those being involved in the January 6th events: those who openly display a military or veteran affiliation, those who identify themselves with the term ``patriot" and those who use QAnon related terms in their names \cite{NPR:2021}. To find these users, we used regex string matching on their ``author names" and their ``author user names". Each user had both an ``author user names" which is his actual account handle, or @-name, and an ``author names" which is a short free-text description the user writes themselves. 
Table \ref{tab:user_def_terms} summarizes the terms used for finding accounts belonging to these three classes of authors. We then analyzed the social media usage and artifacts of these three classes of users. 
Using these three user classes, we analyzed the textual content of these three classes of users and identified coordinated messaging among the user classes.

\begin{table}[]
\begin{tabular}{|l|p{10cm}|}
\hline
\textbf{User Category}    & \textbf{Terms used to define that user category}   \\ \hline
Military/Veteran & \begin{tabular}[c]{@{}l@{}}army, navy, air force, airforce, marine, veteran, military, \\ servicemember, coastguard, coast guard, soldier, infantry, sergeant\end{tabular} \\ \hline
Patriot  & patriot   \\ \hline
QAnon   & qanon, wwg1wga, Q, thegreatawakening, thestorm, theplan  \\ \hline
\end{tabular}
\caption{Terms used to heuristically define three main user classes}
\vspace{-1.3cm}
\label{tab:user_def_terms}
\end{table}

\subsection{Identification of Coordinated Messaging} 
Finally, we use the textual information to evaluate user  coordination in spreading messages themes. 
The Parler data shows a low proportion of parley echoes, which means users are not actively sharing other user's parleys. In contrast, Twitter data has a huge proportion of retweets \cite{hitkul2021capitol}. This renders using methods of identifying user coordination through retweets and co-tweets inadequate, hence we turn to analyzing similar texts.

Similar to previous work, we create user-to-user graphs based on users' textual content to analyze user accounts with very similar textual messages \cite{Pacheco:2020}. However, instead of only creating an edge if there is at least one pair of posts between two users that are above an arbitrary similarity threshold, we use a text-to-text graph to induce a user-to-user graph, and set the threshold of which we filter edges through a statistical analysis of all graph edge weights. 

We consider only original parleys and not echoed-parleys in order to highlight authors who are consciously writing similar texts, but may be trying to hide their affiliation by not explicitly echoing each other. 
We first begin by creating a user-to-text binary graph $P$, which represents the users that wrote each parley. 
Next, we create a text-to-text graph $A$ through a k-Nearest Neighbor (kNN) representation of the parley texts, with $k=log_2{N}$ where $N$ is the number of parleys \cite{Maier:2009}. To do so, we perform BERT vectorization on each parley text create contextualized embeddings into a 768-dimensional latent semantic space \cite{DBLP:journals/corr/abs-1810-04805}. We make use of the FAISS library to index the text vectors and perform an all-pairs cosine similarity search to determine the top k closest vectors to each parley vector \cite{faiss}. We then symmetrize this kNN-graph via $P' = \frac{P+P^{T}}{2}$ to produce a symmetric kNN graph of the posts, as this tends to better maintain meso-structures from the data, like clusters, in the graph \cite{Maier:2011}, \cite{Ruan:2009}, \cite{Campedelli:2019}. The edges of the graph are weighted between $[0,1]$ by the cosine similarity of the two parleys.

Having found a latent graph of the textual content of the posts we then induce a user-to-user graph. We do this through matrix Cartesian product formula of:

\begin{equation}
    U = PA'P^{T}
\end{equation}

Where $U$ is the user-to-user graph, $A$ is a user-to-text bipartite graph where an edge indicates that a user posted a given parley, and $P'$ is the text-to-text kNN graph as previously defined. The resulting graph, $U$ has edges that represent the strength of textual similarity between two users, given how close in similarity their posts are in a latent semantic space. $U$ better accounts for not only having multiple, similar posts but better respect the textual data manifold when measuring between two users as compared at only identifying the most similar posts between two users \cite{Pacheco:2020}.

To sieve out the core structure of the graph, we further prune the graph $U$ based on link weights, forming $U'$, keeping only the links that weight greater than one standard deviation away from the mean link weight. This leaves behind users that strongly resemble each other in terms of the semantic similarity of their parley texts.

\section{Results}

In this section, we detail the results of our analysis. We begin with presenting the social media usage statistics of the three classes of users. We then look at the textual content being spread for disinformation narratives. Finally, we present the results of analyzing users and textual content.

\subsection{Social Media Usage and Artifacts}

After having identified classes of users (i.e. military/veteran, patriot, QAnon) using regex searches within their author names, we then analyzed the differences in their activity on Parler. For the military/veteran users we identified 975 accounts (0.3\% of all accounts) that posted 14,745 Parleys (0.8\% of all Parleys). For the patriot users we identified 5,472 accounts (1.9\% of all accounts) that posted 81,293 Parleys (4.7\% of all Parleys). For the QAnon users we identified 3,400 accounts (1.2\% of all accounts) that posted 71,998 Parleys (4.1\% of all Parleys). So, while these accounts make up a small portion of the users, less than 5\% of the all of the user accounts, they account for close to 10\% of all of the posts by volume. Table \ref{tab:social_media_artifacts} summarizes the social media engagement and use of various social media artifacts across all of the accounts. 

{\footnotesize
\begin{longtable}{|l|llll|}
\hline
                                                                             & All Users                                                   & \begin{tabular}[c]{@{}l@{}}Military/\\ Veteran\end{tabular} & Patriot                                                    & QAnon                                                      \\ \hline
Avg. echos by users                                                          & 6.33+/-15.49                                                & 2.05+/-6.92                                                 & 2.65+/-6.96                                                & 2.65+/-6.88                                                \\ \hline
Avg. users echoed                                                            & 13.35+/-736.96                                              & 18.81+/-203.26                                              & 20.31+/-242.05                                             & 38.66+/-518.36                                             \\ \hline
\begin{tabular}[c]{@{}l@{}}Avg. posts per user with\\ a mention\end{tabular} & 0.09+/-0.8                                                  & 0.16+/-0.9                                                  & 0.10+/-0.49                                                & 0.16+/-0.86                                                \\ \hline
\begin{tabular}[c]{@{}l@{}}Avg. posts per user with\\ a hashtag\end{tabular} & 0.20+/-1.14                                                 & 0.24+/-0.84                                                 & 0.34+/-1.27                                                & 0.34+/-1.2                                                 \\ \hline
\begin{tabular}[c]{@{}l@{}}Avg. posts per user with\\ a URL\end{tabular}     & 0.73+/-2.83                                                 & 0.87+/-2.46                                                 & 0.91+/-2.89                                                & 1.18+/-4.64                                                \\ \hline
\begin{tabular}[c]{@{}l@{}}Avg. posts per user with\\ media\end{tabular}     & 0.47+/-2.02                                                 & 0.88+/-2.77                                                 & 0.62+/-1.61                                                & 1.02+/-3.24                                                \\ \hline
\begin{tabular}[c]{@{}l@{}}Avg. comments received\\ per post\end{tabular}    & 4.28+/-195.77                                               & 3.66+/-26.24                                                & 3.07+/-24.27                                               & 6.96+/-103.20                                              \\ \hline
\begin{tabular}[c]{@{}l@{}}Avg. upvotes received\\ per post\end{tabular}     & 29.53+/-1416.6                                              & 28.57+/-213.84                                              & 32.81+/-329.84                                             & 67.93+/-874.82                                             \\ \hline
\begin{tabular}[c]{@{}l@{}}Avg. impressions received\\ per post\end{tabular} & \begin{tabular}[c]{@{}l@{}}4312.42+/-\\ 241116\end{tabular} & \begin{tabular}[c]{@{}l@{}}3946.72+/-\\ 36358\end{tabular}  & \begin{tabular}[c]{@{}l@{}}3002.06+/-\\ 29649\end{tabular} & \begin{tabular}[c]{@{}l@{}}6666.75+/-\\ 76551\end{tabular} \\ \hline
\begin{tabular}[c]{@{}l@{}}Mode impressions received\\ per post\end{tabular} & 10                                                          & 11                                                          & 24                                                         & 12                                                         \\ \hline
\caption{Social media usage patterns across the three user classes. Most users have very few echos and use very few (or none) social media artifacts, like hashtags or mentions. Users that identify with QAnon terms use more social media artifacts and see greater engagement.}
\label{tab:social_media_artifacts}
% \vspace{-0.4cm}
\end{longtable}
}

Generally, we observe that the social engagement is different for the accounts that identify with QAnon and patriot terms, as compared to the overall Parler population. The military, patriot and QAnon accounts are all echoed more than the population average. This trend of having posts be highly echoed is especially true for the QAnon accounts which, on average, see three times as much echoing on their posts than the average user does. Additionally, the QAnon user category also sees more impressions, upvotes, and comments than does any other category of user. The military, patriot, and QAnon accounts also use more social media artifacts, like hashtags or mentions than do the average account. So, those accounts that openly identify with QAnon terms tend to see more social engagement with their posts than any other class of accounts and all of the special types of accounts use more social media artifacts than average users.

Additionally, the verbiage used by these accounts tends toward those associated with disinformation campaigns. The top words by their counts within the textual content of all the posts of all three accounts includes the following words: trump, pence, patriot, people, antifa, country, president, dc, echo, election, need, today, wwg1wga, and stopthesteal. While some of these terms are common throughout the data set, like `trump', `need', `president', or `people', others show up more prominently in the three classes of accounts, like `wwg1wga' and `antifa', than they do across all posts. These usage of these words are almost uniquely associated with disinformation campaigns like the 2020 U.S. Election being fraudulent or Antifa being responsible for the January 6th Capitol Riot. Thus, we observe that the three classes of users also frequently employ verbiage associated with disinformation.

\subsection{Coordinated Messaging between Users}
\begin{figure}[!htb]
    \centering
    \includegraphics[width=1.0\textwidth]{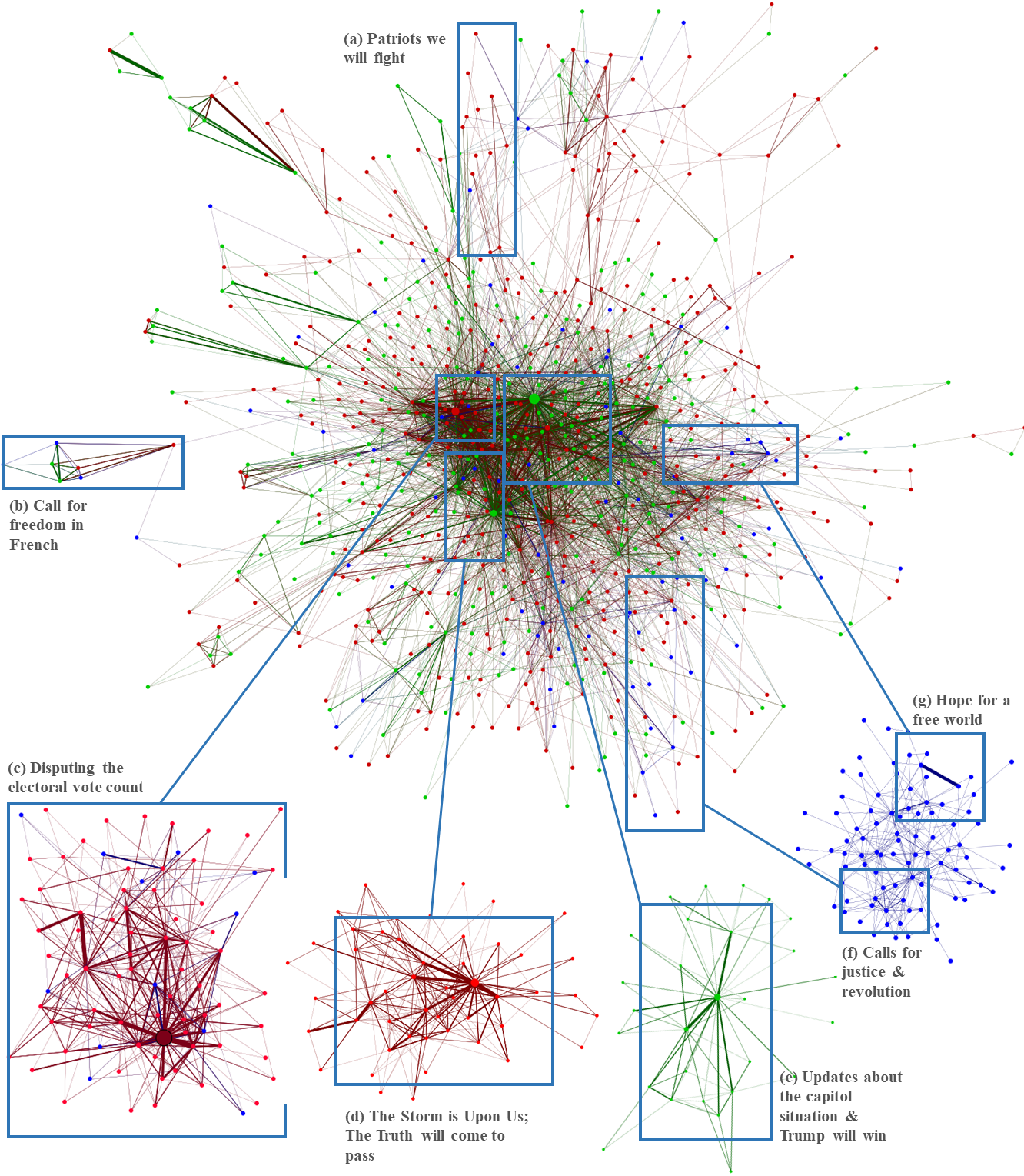}
    \caption{Core structure of user-to-user graph $U$' representing textual coordination between groups of users. Military users are colored blue; patriot users red and QAnon users green. The thickness of the links represent the strength of the coordination. Nodes are sized by total degree centrality value.}
    \label{fig:overallresults}
\vspace{-0.5cm}
\end{figure}

Figure \ref{fig:overallresults} presents the graph $U'$, the core structure of the induced user-to-user graph $U$. We first observe that this graph has a distinct core structure which is due to the fact that almost all of the conversation surrounds the 2020 U.S. Election. This graph is annotated with clusters of coordinated messaging between users derived from the analysis of Parley textual similarity, where the strength of the coordination is represented by the link weights. We segment the graph $U'$ into subgraphs (a) to (g) for better clarity of each portion of the graph. The messaging of each subgraph are consistent, showing the effectiveness of our approach in uncovering coordination between users as well as distinct disinformation themes spread by those users.

Subgraph (a) represents Patriot users who have every conviction they will fight for their Republic and president. 
Subgraph (b) calls for freedom in French, with many users having handles that end with the different Canadian provinces. The users associate themselves with different user groups.
Subgraph (c) sieves out users in the center of the graph that strongly coordinate with each other in disputing the electoral vote count. These are predominantly patriot users and a handful of military users disputing the vote.

Subgraph (d) in which the core patriot in the center of the graph strongly coordinates with each other. This user group mainly advocates that the truth about the elections will come to pass and the storm is upon us. These messages express support for the disinformation narratives of the 2020 U.S. Election being fraudulent as well as QAnon conspiracy theories relating to then President Donald Trump seizing power in the U.S. 

Subgraph (e) represents QAnon users, who mostly purport that Trump will win, and provide updates about the capitol situation in terms of spurring the group on (``PLS SPREAD TrumpVictory FightBack ProtectPatriots") and information updates (``Protesters are now in Nancy Pelosi‘s office"). Some of these narratives are blaming BLM and a deliberately designed terrorist Antifa who infiltrated protesters who stormed the Capitol.

Subgraph (f) sieves out users at the fringe of the graph that call for justice and revolution and not to be let down. 
Subgraph (g) represents military users that hope for a free world and generally call to Stop the Steal of the 2020 elections. ``Stop the Steal" has been a recurring phrase throughout Trump's political career, peaking in the Capitol riots as a movement to overturn the 2020 US election results and has been a key phrase associated with the disinformation campaign trying to de-legitimize the 2020 U.S. election.

\section{Discussion and Conclusion} 

In this work, we analyzed actors involved in disinformation spread on Parler during the January 6th event. We looked at three main user groups based upon unusual online activity and their salience in stories associated with the January 6th riots: users that openly present a military or veteran affiliation, user that use the moniker `patriot', and users that identify with QAnon-related terms. We found that the posts of these three groups of users are echoed more and receive more impressions than the population average, with the QAnon users having the most of all. These user groups also employ typical tactics of using more hashtags and mentions than the average accounts to spread their messages, as well as key verbiage relating to disinformation campaigns. So, we find that certain users that self-identify with particular identities typically revered by the politically right in the U.S. (i.e. military service, patriotism, etc.) tend be more involved in guiding the conversation within right-leaning social media spheres.

Having identified these influential accounts, we then looked at textual coordination between actors. Identifying the strength of user-to-user coordination over link weights presents coordinated activity as a spectrum rather than a binary state of shared activity \cite{vargas2020detection}. This shows which users typically post similar texts, allow identification of how similar user accounts are to each other.  Our approach of inducing a user-to-user graph by their textual similarity produced distinct clusters of users that had similar textual themes. These themes include known disinformation themes like the dispute of the electoral vote count and that Trump will win the 2020 US elections. So, within Parler, we observed distinct groups of users, typically that identify as patriot or QAnon that did engage in posting the same textual content, that aligns to known 2020 election narratives.

There are some important limitations to this study. First, the data collected was only a partial scrape of the parleys made on the platform surrounding the Capitol riots event. As the platform was shut down, we cannot conclude this was the entire set of parleys related to the event. Nonetheless, we believe that the huge volume of parleys should provide sufficient generalizability of the online discourse around this event. Second, we have not explored all of the means of properly learning the textual similarity graph or sieving out the main graph structure from the induced graph. So, we do not know if we have found the optimal graph representation.

Future work include looking at a multimodal coordination effort on the platform in contributing to the spread of disinformation, rather than only the text field, taking into account the multitude of images and videos posted on Parler, resulting in the offline violence. Another important future direction is to study dynamic coordination networks, based upon temporal information to characterize the change in coordinated messaging among users. 

In this study we used a general approach to sieve out the core structure of a user-to-user coordination graph which relies textual content of a social media data. This provides a more robust filtering approach as compared to work that relies on an arbitrary link weight threshold value, below which the links are discarded. Our approach can further help to identify disinformation campaigns where automated groups of user accounts post a consistent similar message, and can overcome shortcomings with just trying to find coordination among disinformation actors by their explicit networks (i.e. echoing on Parler). We hope the techniques here can be used to better analyze social media platforms and streamline network structures into core components for further investigation. 
The early detection of coordinated effort to organize riots and political violence online may present the possibility of stopping that violence in the real world.

\bibliographystyle{splncs04}
\bibliography{biblography}
\end{document}